\documentclass[10pt, a4paper]{article}

\usepackage[english]{babel}

\usepackage[
backend=biber,
style=alphabetic,
sorting=ynt
]{biblatex}
\usepackage[letterpaper,top=2cm,bottom=2cm,left=3cm,right=3cm,marginparwidth=1.75cm]{geometry}

\usepackage{amsmath}
\usepackage{amssymb,bm, amsbsy}
\usepackage{graphicx}
\usepackage{calligra}
\usepackage{authblk}
\usepackage{csquotes}

\newcommand{\CC}{\mbox{\calligra{\scriptsize{C}}}}
\newcommand{\C}{\bm{\mathsf{C}}}
\newcommand{\PP}{\bm{\mathsf{P}}}
\newcommand{\TT}{\bm{\mathsf{T}}}
\newcommand{\HH}{\bm{\mathsf{H}}}
\newcommand{\E}{\bm {\varepsilon}}
\newcommand{\ftilde}{\raisebox{.7em}{\resizebox{.6em}{!}{\ensuremath{\bm{\sim}}}}\hspace{-.7em}}

\newcommand{\ols}[1]{\mskip.5\thinmuskip\overline{\mskip-.5\thinmuskip {#1} \mskip-.5\thinmuskip}\mskip.5\thinmuskip} 
\newcommand{\olsi}[1]{\,\overline{\!{#1}}} 
\makeatletter
\newcommand\closure[1]{
  \tctestifnum{\count@stringtoks{#1}>1} 
  {\ols{#1}} 
  {\olsi{#1}} 
}
\long\def\count@stringtoks#1{\tc@earg\count@toks{\string#1}}
\long\def\count@toks#1{\the\numexpr-1\count@@toks#1.\tc@endcnt}
\long\def\count@@toks#1#2\tc@endcnt{+1\tc@ifempty{#2}{\relax}{\count@@toks#2\tc@endcnt}}
\def\tc@ifempty#1{\tc@testxifx{\expandafter\relax\detokenize{#1}\relax}}
\long\def\tc@earg#1#2{\expandafter#1\expandafter{#2}}
\long\def\tctestifnum#1{\tctestifcon{\ifnum#1\relax}}
\long\def\tctestifcon#1{#1\expandafter\tc@exfirst\else\expandafter\tc@exsecond\fi}
\long\def\tc@testxifx{\tc@earg\tctestifx}
\long\def\tctestifx#1{\tctestifcon{\ifx#1}}
\long\def\tc@exfirst#1#2{#1}
\long\def\tc@exsecond#1#2{#2}
\makeatother
\usepackage[T1]{fontenc}
\usepackage[colorlinks=true, allcolors=blue]{hyperref}

\title{12-dimensional Lie Algebra of Entangled Spin Fields}
\author{Alexandru Gabriel Mitru\c{t} \thanks{amitrut@gmail.com}}
\affil{Universitatea din Bucure\c{s}ti}
\date{April 2, 2024}
\begin{document}
\maketitle
\begin{abstract}
In this paper was proved (i) the equivalence of Dirac equation for mass $\mathsf{m}$ with two entangled Proca fields of mass $\mathsf{2m}$ and (ii) is proposed an equation and (iii)  an 12-dimensional Lie algebra for the entangled spin fields. 
\end{abstract}
\section{Introduction}
In addition to the well known Dirac, Weyl and Majorana representations of gamma matrices, we define a new one as 
\begin{align}
  \gamma^{0}&=\begin{pmatrix}
0 & i & 0 & 0 \\
-i & 0 & 0 & 0\\
0 & 0 & 0 & 1\\
0 & 0 & 1 & 0
\end{pmatrix},\ &
\gamma^{3}&=\begin{pmatrix}
0 & i & 0 & 0 \\
i & 0 & 0 & 0\\
0 & 0 & 0 & 1\\
0 & 0 & -1 & 0
\end{pmatrix},\ 
  \end{align}
  \begin{align}
  \gamma^{1}&=\begin{pmatrix}
0 & 0 & 1 & 0 \\
0 & 0 & 0 & i\\
-1 & 0 & 0 & 0\\
0 & i & 0 & 0
\end{pmatrix},\ &
\gamma^{2}&=\begin{pmatrix}
0 & 0 & i & 0 \\
0 & 0 & 0 & -1\\
i & 0 & 0 & 0\\
0 & 1 & 0 & 0
\end{pmatrix}.
  \end{align}
Next we solve Dirac equation to get two plane wave solutions $\Phi_{1}(p)$ and $\Phi_{2}(p)$ of positive energy,
\begin{align}
\begin{aligned}
\HH \Phi_{1}(p) & =+\E\,\Phi_{1}(p)\quad,\
\end{aligned}
&&
\begin{aligned}
\HH \Phi_{2}(p)& =+\E\,\Phi_{2}(p)\quad,\
\end{aligned}
  \end{align}
and two plane wave solutions $\Psi_{1}(p)$ and $\Psi_{2}(p)$ of negative energy,
\begin{align}
\begin{aligned}
\HH \Psi_{1}(p) & =-\E\,\Psi_{1}(p)\quad,\
\end{aligned}
&&
\begin{aligned}
\HH \Psi_{2}(p)& =-\E\,\Psi_{2}(p)\quad,\
\end{aligned}
  \end{align}
where \,$\HH=\gamma^{0}(-i\gamma^{i}\partial_{i}+m)\,$denotes the Hamiltonian, and $\E=+(\mathbf{p}^2+m^2)^{1/2}$. Explicitly, the plane wave solutions in this representation are

\begin{align}\label{positive-solutions}
  \begin{aligned}
  \Phi_{1}(p) & =
  \frac{1}{\left[2 \E (\E + p_{3})\right]^{1/2}}
\begin{pmatrix}
m\\-i(\E+p_{3})\\p_{1}-i p_{2}\\0
\end{pmatrix} e^{-i p x}\quad,\
\end{aligned}
  &&
  \begin{aligned}
  \Phi_{2}(p) & =\frac{1}{\left[2 \E (\E - p_{3})\right]^{1/2}}
\begin{pmatrix}0\\
p_{2}-i p_{1}\\
\E-p_{3}\\
m
\end{pmatrix} e^{-i p x}\quad,\
  \end{aligned}
  \end{align}
  and 
\begin{align}\label{negative-solutions}
  \begin{aligned}
  \Psi_{1}(p)& =
  \frac{1}{\left[2 \E (\E + p_{3})\right]^{1/2}}
\begin{pmatrix} p_{1}+i p_{2} \\
0\\
m\\
-\E-p_{3}\end{pmatrix} e^{+i p x}\quad,\
  \end{aligned}
  &&
  \begin{aligned}
  \Psi_{2}(p)&=\frac{1}{\left[2 \E (\E - p_{3})\right]^{1/2}}\begin{pmatrix}
-i(\E-p_{3})\\
m\\
0\\
p_{2}+i p_{1}
\end{pmatrix} e^{+i p x}\quad.\
  \end{aligned}
  \end{align}
\section{Parity, Chirality and Charge Conjugation Operators}
Alongside the chirality operator $\gamma_{5}$,  we define $\PP=-i\gamma^{1}\gamma^{2}$ the parity operator and $\C$ charge (parity and chirality) conjugation operator as

   \begin{align}
    \PP&=\begin{pmatrix}
    1 & 0 & 0 & 0\\
    0 & 1 & 0 & 0\\
    0 & 0 & -1 & 0\\
    0 & 0 & 0 & -1\\
    \end{pmatrix},\ &
    \gamma_{5}&=\begin{pmatrix}
    1 & 0 & 0 & 0\\
    0 & -1 & 0 & 0\\
    0 & 0 & -1 & 0\\
    0 & 0 & 0 & 1\\
    \end{pmatrix},\ &
     \C&=\begin{pmatrix}
    0 & 0 & 1 & 0\\
    0 & 0 & 0 & i\\
    1 & 0 & 0 & 0\\
    0 & i & 0 & 0\\
    \end{pmatrix}\CC
    \end{align}
where $\gamma_{5}=i\gamma^{0}\gamma^{1}\gamma^{2}\gamma^{3}$ and $\CC$ \, is the complex conjugation operator. Chirality and parity conjugation operator $\C$ is  antilinear and anticommutes with the Hamiltonian,
\begin{align}
  \begin{aligned}
  \C^{2}&=1\quad,\
  \end{aligned}
   \begin{aligned}
  \HH\C=-\C\HH\quad.\
  \end{aligned}
  \end{align}
therefore if $\Psi$ is a positive(negative) energy solution then $\Phi=\C\Psi$ is a negative (positive) solution. While charge conjugation operators transform a spinor into another of opposite energy leaving its four-momentum unchanged, $\PP$ transform a spinor with four-momentum $p^{\alpha}=(\mathbf{\varepsilon}, p_{1},p_{2},p_3)$ into the same spinor but with four-momentum $\gamma_{5}\,p=(\mathbf{\varepsilon}, -p_{1},-p_{2},p_3)$. The commutation relation $\bm{[}\,\gamma_{5}\,,\,\PP\,\bm{]}=0$ imply that eigenvalues of chirality $\gamma_{5}$ and parity $\PP$ can be used to classify their common eigenstates. The $-1$ and $+1$ eigenvalues of $\gamma_{5}$ will be denoted $\mathsf{L}$ and $\mathsf{R}$ i.e. $-1$ correspond to left chiral $\mathsf{L}$ and $+1$ to right chiral $\mathsf{R}$, while the eigenvalues $-1$ and $+1$ of $\PP$ will be denoted as $-$ and $+$ i.e. $-$ for negative parity eigenvalue $-1$ and $+$ for positive parity $+1$. We define the eight common eigenstates of chirality $\gamma_{5}$ and parity $\PP$ using the solutions $\Phi$ and $\Psi$ of opposite energy 
\begin{align}\label{proto}
  \begin{aligned}
        \Psi_{\pm\mathsf{L}}&=\frac{1\pm\PP}{2}\frac{1-\mathsf{\gamma_{5}}}{2}\Phi\qquad,\\\
         \Phi_{\pm\mathsf{L}}&=\frac{1\pm\PP}{2}\frac{1-\mathsf{\gamma_{5}}}{2}\Psi\qquad,\
    \end{aligned}
    &&
     \begin{aligned}
         \Psi_{\pm\mathsf{R}}&=\frac{1\pm\PP}{2}\frac{1+\mathsf{\gamma_{5}}}{2}\Phi\qquad,\\\
         \Phi_{\pm\mathsf{R}}&=\frac{1\pm\PP}{2}\frac{1+\mathsf{\gamma_{5}}}{2}\Psi\qquad,\
    \end{aligned}
\end{align}
Charge (chirality and parity) conjugation operator $\C$ transform  $\Phi$ eigenstates of chirality and parity into  $\Psi$ eigenstates with inverse eigenvalues of chirality and parity, and vice-versa  
\begin{align}\label{aproto}
  \begin{aligned}
     \Psi_{+\mathsf{L}}&=\C \Phi_{-\mathsf{R}}\qquad,\\\
         \Psi_{-\mathsf{L}}&=\C \Phi_{+\mathsf{R}}\qquad,\\\
          \Psi_{+\mathsf{R}}&=\C \Phi_{-\mathsf{L}}\qquad,\\\
           \Psi_{-\mathsf{R}}&=\C \Phi_{+\mathsf{L}}\qquad,\
  \end{aligned}
  &&
    \begin{aligned}
     \Phi_{-\mathsf{R}}&=\C \Psi_{+\mathsf{L}}\qquad,\\\
         \Phi_{+\mathsf{R}}&=\C \Psi_{-\mathsf{L}}\qquad,\\\
          \Phi_{-\mathsf{L}}&=\C \Psi_{+\mathsf{R}}\qquad,\\\
           \Phi_{+\mathsf{L}}&=\C \Psi_{-\mathsf{R}}\qquad.\
  \end{aligned}
  \end{align}
Next, we define left $X_{\mathsf{L}}$ and right $X_{\mathsf{R}}$ eigenstates of chiral operator  by superposition of all left chiral and respectively right  eigenstates
\begin{align}\label{condensation}
    \begin{aligned}
    X_{\mathsf{L}}&=\Phi_{\mathsf{-L}}+\Phi_{\mathsf{+L}}+\Psi_{\mathsf{-L}}+\Psi_{\mathsf{+L}}\quad,
    \end{aligned}
    &&
    \begin{aligned}
      X_{\mathsf{R}}&=\Phi_{\mathsf{-R}}+\Phi_{\mathsf{+R}}+\Psi_{\mathsf{-R}}+\Psi_{\mathsf{+R}}\quad,
    \end{aligned}
\end{align}
which form an orthogonal set  
\begin{align}
 \begin{aligned}
 \closure{X}_{\mathsf{L}}X_{\mathsf{L}}&=0\quad,\\\
 \closure{X}_{\mathsf{L}}X_{\mathsf{R}}&=0\quad,\
    \end{aligned}
    &&
    \begin{aligned}
    \closure{X}_{\mathsf{R}}X_{\mathsf{R}}&=0\quad,\\\
    \closure{X}_{\mathsf{R}}X_{\mathsf{L}}&=0\quad,\
    \end{aligned}
\end{align}
 mix positive (negative) energy solutions with negative (positive) energy solutions of Dirac equation
\begin{align}
\begin{aligned}
X_{\mathsf{L}}&=\frac{1-\gamma_{5}}{2}(\Psi+\Phi)\quad,\\\
  \closure X_{\mathsf{L}}&=(\closure\Psi+\closure\Phi)\frac{1+\gamma_{5}}{2}\quad,\
\end{aligned}
&&
\begin{aligned}
X_{\mathsf{R}}&=\frac{1+\gamma_{5}}{2}(\Psi+\Phi)\quad,\\\
    \closure X_{\mathsf{R}}&=(\closure\Psi+\closure\Phi)\frac{1-\gamma_{5}}{2}\quad,\
\end{aligned}
\end{align}
which satisfy Proca equation
\begin{align}
    \begin{aligned}
    (\square+m^2)X_{\mathsf{L}}=0\quad,\
    \end{aligned}
    &&
    \begin{aligned}
    (\square+m^2)X_{\mathsf{R}}=0\quad,\
    \end{aligned}
\end{align}
and the following partial differential coupled equations
\begin{align}
    \begin{aligned}
    i\,\partial_{\alpha}\gamma^{\alpha}X_{\mathsf{L}}-mX_{\mathsf{R}}&=0\quad,\
    \end{aligned}
    &&
    \begin{aligned}
    i\,\partial_{\alpha}\gamma^{\alpha}X_{\mathsf{R}}-mX_{\mathsf{L}}&=0\quad.\
    \end{aligned}
\end{align}
\section{Conserved four-currents}
 We introduce left and right conserved four-currents (probability densities) for eigenstates of the same chirality defined by \ref{condensation}
\begin{align}
    \begin{aligned}
       j^{\alpha}({X_{\mathsf{L}}},X_{\mathsf{L}})&=\closure{X}_{\mathsf{L}}\gamma^{\alpha}X_{\mathsf{L}}\quad,\
    \end{aligned}
    &&
     \begin{aligned}
        j^{\alpha}({X_{\mathsf{R}}},X_{\mathsf{R}})&=\closure{X}_{\mathsf{R}}\gamma^{\alpha}X_{\mathsf{R}}\quad,\
    \end{aligned}
\end{align}
which are equal and real defined
\begin{align}
\begin{aligned}
\CC\,\,j^{\alpha}(X_{\mathsf{L}},X_{\mathsf{L}})&=j^{\alpha}(X_{\mathsf{L}},X_{\mathsf{L}})\quad,\
\end{aligned}
&&
\begin{aligned}
\CC\,\,j^{\alpha}(X_{\mathsf{R}},X_{\mathsf{R}})&=j^{\alpha}(X_{\mathsf{R}},X_{\mathsf{R}})\quad.\
\end{aligned}
\end{align}
We interpret the conservation of the left and right four-currrent
\begin{align}\label{LR-conservation}
    \begin{aligned}
  \partial_{\alpha}\,j^{\alpha}({X_{\mathsf{L}}},X_{\mathsf{L}})&=0\quad,\
    \end{aligned}
    &&
      \begin{aligned}
    \partial_{\alpha}\,j^{\alpha}({X_{\mathsf{R}}},X_{\mathsf{R}})&=0\quad,\
    \end{aligned}
\end{align}
and the zero current between eigenstates of left an right chirality
\begin{equation}
j^{\alpha}({X_{\mathsf{L}}},X_{\mathsf{R}})=j^{\alpha}({X_{\mathsf{R}}},X_{\mathsf{L}})=0\quad,\
\end{equation}
as the existence of chiral particles $\mathsf{L}$ and $\mathsf{R}$. The two chiral particles are transformed one into another by charge ( chirality and parity) operator $\C$ \begin{align}
\begin{aligned}
X_{\mathsf{L}}=\C X_{\mathsf{R}}\quad,\
\end{aligned}
&&
\begin{aligned}
X_{\mathsf{R}}=\C X_{\mathsf{L}}\quad,\
\end{aligned}
\end{align}
 The particle described by $\Psi+\Phi=X_{\mathsf{L}}+X_{\mathsf{R}}$, which has no definite chirality nor parity and is an eigenstate of chirality and conjugation operator, i.e. is transformed into itself by chirality and parity operator,
\begin{equation}
\C(\Psi+\Phi)=\Psi+\Phi\quad,\  
\end{equation}
therefore it is a Majorana particle. All three particles mix positive (negative) energy solutions of Dirac equation with negative (positive) energy solutions, and their currents are explicitly given by
\begin{align}\label{LR}
    \begin{aligned}
    j^{\alpha}(X_{\mathsf{L}},X_{\mathsf{L}})&=(\closure{\Psi}+\closure{\Phi})\gamma^{\alpha}\frac{1-\gamma_{5}}{2}(\Psi+\Phi)\quad,\
    \end{aligned}
    &&
    \begin{aligned}
    j^{\alpha}(X_{\mathsf{R}},X_{\mathsf{R}})&=(\closure{\Psi}+\closure{\Phi})\gamma^{\alpha}\frac{1+\gamma_{5}}{2}(\Psi+\Phi)\quad,\
    \end{aligned}
\end{align}
and
\begin{align}
    \begin{aligned}
    j^{\alpha}(\Psi+\Phi,\Psi+\Phi)=(\closure{\Psi}+\closure{\Phi})\gamma^{\alpha}(\Psi+\Phi)\quad.\
    \end{aligned}
\end{align}
Out of 64 currents between all 8 common eigenstates of chirality and parity defined by \ref{proto}, 32 of them,  between $\mathsf{L}$ and $\mathsf{R}$ or $\mathsf{R}$ and $\mathsf{L}$ are zero. The 32 non-zero currents are between eigenstates of the same chirality, but are not conserved.
\section{Parity Violation}
Next, we calculate in the ultrarelativistic regime $\E\gg m$ the helicity of $\mathsf{L}$ and $\mathsf{R}$ particles moving along the $z$ axis. The helicity operator is given by
\begin{equation}
    \bm{\mathsf{h}}=\frac{1}{p}(p_{1}S^{1}+p_{2}S^{2}+p_{3}S^{3})\quad,\
\end{equation}
where $p=(p_{1}^{2}+p_{2}^{2}+p_{3}^{2})^{1/2}$ is momentum and  $S^{1}=\Sigma^{23}$, $S^{2}=\Sigma^{31}$, $S^{3}=\Sigma^{12}$ are spatial components of spin, related to the generators of Lorentz transformations of gamma matrices by $\Sigma^{\mu\nu}=\frac{i}{4}[\gamma^{\mu},\gamma^{\nu}]$. Helicity commutes with parity and chirality operator and for a particle moving along $z$ axis we get $\bm{\mathsf{h}}=-\frac{1}{2}\PP$. For each of the four solution of Dirac equations we calculate the action of helicity operator on ultrarelativistic left and right particles moving  along $z$ axis and get correct predictions
\begin{align}
    \begin{aligned}
    \bm{\mathsf{h}}X_{\mathsf{L}}&=-\frac{1}{2} X_{\mathsf{L}}\quad,\
    \end{aligned}
    &&
    \begin{aligned}
    \bm{\mathsf{h}}X_{\mathsf{R}}&=+\frac{1}{2} X_{\mathsf{R}}\quad,\
    \end{aligned}
\end{align}
when $\E \gg m$. We can now identify the neutrino as the left chiral particle $X_{\mathsf{L}}$ having helicity $-\frac{1}{2}$ and the antineutrino as the right chiral particle $X_{\mathsf{R}}$ having helicity $+\frac{1}{2}$. Conserved currents  are formed only by left $X_{\mathsf{L}}$ and  right $X_{\mathsf{R}}$ eigenstates of chirality, not by the parity eigenstates. The superposition of all negative $\Psi_{-\mathsf{L}}+\Psi_{-\mathsf{R}}+\Phi_{-\mathsf{L}}+\Phi_{-\mathsf{R}}$ parity eigenstates
and of all positive parity eigenstates $\Psi_{+\mathsf{L}}+\Psi_{+\mathsf{R}}+\Phi_{+\mathsf{L}}+\Phi_{+\mathsf{R}} $
 gives currents which are are not conserved, i.e. we can not built conserved currents using linear combinations of parity eigenstates. Therefore, it is the chirality which is the fundamental symmetry of nature, because the left chiral and right chiral currents are conserved and identified with $\mathsf{L}$ and right $\mathsf{R}$ particle, i.e. neutrino and respectively antineutrino and the question is if chirality, not parity, is violated, i.e. if there is a field (particle) which couple with different strengths to $\mathsf{L}$ and $\mathsf{R}$ currents. The property of parity operator, in Weyl basis, to transform the left-handed Weyl spinor into the right-handed Weyl spinor and vice-versa requires that  right-handed neutrino and left-handed antineutrino should also exist. The absence of right-handed neutrino and left-handed antineutrino had led to the conclusion that parity is not conserved. The mistake was to associate left-handed Weyl spinor, which gives a four current that is not conserved, to neutrino, and similarly, a right-handed Weyl spinor, which gives a four current that is not conserved, to antineutrino. Our definition of  currents associated to left and right eigenstates of chirality are conserved \ref{LR-conservation} and transformed one into another by the chirality and parity conjugation operator \ref{aproto}, not by the parity operator as in Weyl representation. Therefore, the puzzling absence of positive helicity neutrino and respectively negative helicity antineutrino, dubbed as "parity violation" is a consequence wrong association of spinors to particles. The $-1/2$ helicity of  the neutrino and $+1/2$ helicity of antineutrino are a result of their  ultrarelatistic regime in which eigenstates $X_{\mathsf{L}}$ and $X_{\mathsf{R}}$ of chirality operator $\gamma_{5}$ are also eigenstates of parity operator $\PP$.
\section{Entangled spin fields equation}
We construct conserved currents with first order derivatives of left eigenstates $\partial^{\mu}X_{\mathsf{L}}$ and symmetrical conserved currents with first order derivatives of right  eigenstates $\partial^{\mu}X_{\mathsf{R}}$,
\begin{align}
    \begin{aligned}
    j_{(1)}^{\alpha}(X_{\mathsf{L}}, X_{\mathsf{L}})&=\partial_{\mu}\closure{X}_{\mathsf{L}}\gamma^{\alpha}\,\partial^{\mu}X_{\mathsf{L}}\quad,\\\
    \partial_{\alpha}\,j_{(1)}^{\alpha}({X_{\mathsf{L}}},X_{\mathsf{L}})&=0\quad,\
    \end{aligned}
    &&
    \begin{aligned}
    j_{(1)}^{\alpha}(X_{\mathsf{R}}, X_{\mathsf{R}})&=\partial_{\mu}\closure{X}_{\mathsf{R}}\gamma^{\alpha}\,\partial^{\mu}X_{\mathsf{R}}\quad,\\\
    \partial_{\alpha}\,j_{(1)}^{\alpha}({X_{\mathsf{R}}},X_{\mathsf{R}})&=0\quad,\
    \end{aligned}
\end{align}
as well as conserved currents with second order derivatives of left and respectively right eigenstates
\begin{align}
    \begin{aligned}
    j_{(2)}^{\alpha}(X_{\mathsf{L}}, X_{\mathsf{L}})&=\partial_{\mu}\partial_{\nu}\closure{X}_{\mathsf{L}}\gamma^{\alpha}\,\partial^{\mu}\partial^{\nu}X_{\mathsf{L}}\quad,\\\
    \partial_{\alpha}\,j_{(2)}^{\alpha}(X_{\mathsf{L}}, X_{\mathsf{L}})&=0\quad,\
    \end{aligned}
    &&
    \begin{aligned}
    j_{(2)}^{\alpha}(X_{\mathsf{R}}, X_{\mathsf{R}})&=\partial_{\mu}\partial_{\nu}\closure{X}_{\mathsf{R}}\gamma^{\alpha}\,\partial^{\mu}\partial^{\nu}X_{\mathsf{R}}\quad,\\\
    \partial_{\alpha}\,j_{(2)}^{\alpha}(X_{\mathsf{R}}, X_{\mathsf{R}})&=0\quad,\
    \end{aligned}
\end{align}
and also conserved currents with derivatives of higher order than 2.
For both left and right particles, only $j^{\alpha}$ and $j^{\alpha}_{(1)}$ are independent, all other currents built with derivatives of higher order than one can be expressed in term of $j^{\alpha}$ or $j^{\alpha}_{(1)}$  as follows
\begin{align}
    \begin{aligned}
    j_{(2)}^{\alpha}(X_{\mathsf{L}}, X_{\mathsf{L}})&=m^{4} j^{\alpha}(X_{\mathsf{L}}, X_{\mathsf{L}})\quad,\\\
    j_{(3)}^{\alpha}(X_{\mathsf{L}}, X_{\mathsf{L}})&=m^{4} j_{(1)}^{\alpha}(X_{\mathsf{L}}, X_{\mathsf{L}})\quad,\
    \end{aligned}
    &&
    \begin{aligned}
    j_{(2)}^{\alpha}(X_{\mathsf{R}}, X_{\mathsf{R}})&=m^{4} j^{\alpha}(X_{\mathsf{R}}, X_{\mathsf{R}})\quad,\\\
    j_{(3)}^{\alpha}(X_{\mathsf{R}}, X_{\mathsf{R}})&=m^{4} j_{(1)}^{\alpha}(X_{\mathsf{R}}, X_{\mathsf{R}})\quad,\
    \end{aligned}
\end{align}
and so on. The currents $j^{\alpha}(X_{\mathsf{L}},X_{\mathsf{L}})$ and $j^{\alpha}_{(1)}(X_{\mathsf{L}},X_{\mathsf{L}})$ satisfy coupled second order differential equations
\begin{align}\label{square}
\begin{aligned}
&\square\,j^{\alpha}(X_{\mathsf{L}},X_{\mathsf{L}})=-2 m^2 j^{\alpha}(X_{\mathsf{L}},X_{\mathsf{L}})+2 j_{(1)}^{\alpha}(X_{\mathsf{L}},X_{\mathsf{L}})\quad,\\\
 &\square\,j_{(1)}^{\alpha}(X_{\mathsf{L}},X_{\mathsf{L}})=-2 m^2 j_{(1)}^{\alpha}(X_{\mathsf{L}},X_{\mathsf{L}})+2 m^4 j^{\alpha}(X_{\mathsf{L}},X_{\mathsf{L}})\quad,\
\end{aligned}
\end{align}
and similar equations holds for $j^{\alpha}(X_{\mathsf{R}},X_{\mathsf{R}})$ and $j^{\alpha}_{(1)}(X_{\mathsf{R}},X_{\mathsf{R}})$. We define four vector potentials $A_{\mathsf{L}}^{\alpha}$, $A_{R}^{\alpha}$, $B_{\mathsf{L}}^{\alpha}$, and $B_{\mathsf{R}}^{\alpha}$
\begin{align}
    \begin{aligned}
    A_{\mathsf{L}}^{\alpha}&=\frac{1}{2}\left [j^{\alpha}(X_{\mathsf{L}},X_{\mathsf{L}})-m^{-2}\,j_{(1)}^{\alpha}(X_{\mathsf{L}},X_{\mathsf{L}})\right]\quad,\\\
    B_{\mathsf{L}}^{\alpha}&=\frac{1}{2}\left [ j^{\alpha}(X_{\mathsf{L}},X_{\mathsf{L}})+m^{-2}\,j_{(1)}^{\alpha}(X_{\mathsf{L}},X_{\mathsf{L}})\right]\quad,\
    \end{aligned}
    &&
     \begin{aligned}
    A_{\mathsf{R}}^{\alpha}&=
    \frac{1}{2}\left [j^{\alpha}(X_{\mathsf{R}},X_{\mathsf{R}})-m^{-2}\,j_{(1)}^{\alpha}(X_{\mathsf{R}},X_{\mathsf{R}})\right]\quad,\\\
    B_{\mathsf{R}}^{\alpha}&=\frac{1}{2}\left [j^{\alpha}(X_{\mathsf{R}},X_{\mathsf{R}})+m^{-2}\,j_{(1)}^{\alpha}(X_{\mathsf{R}},X_{\mathsf{R}})\right]\quad.\
    \end{aligned}
\end{align}
Using \ref{square} find that $A_{\mathsf{L}}^{\alpha}$ and $A_{\mathsf{R}}^{\alpha}$ fields satisfy Proca equation for a particle with mass $2m$ and the Lorenz gauge is automatically held
\begin{align}
    \begin{aligned}
    (\square+4m^2)A_{\mathsf{L}}^{\alpha}&=0\quad,\\\
    \partial_{\alpha}A_{\mathsf{L}}^{\alpha}&=0\quad,\
    \end{aligned}
    &&
    \begin{aligned}
    (\square+4m^2)A_{\mathsf{R}}^{\alpha}&=0\quad,\\\
    \partial_{\alpha}A_{\mathsf{R}}^{\alpha}&=0\quad,\
    \end{aligned}
\end{align}
while $B_{L}^{\alpha}$ and $B_{R}^{\alpha}$ satisfy Maxwell equations
\begin{align}
    \begin{aligned}
    \square B_{\mathsf{L}}^{\alpha}&=0\quad,\\\
    \partial_{\alpha} B_{\mathsf{L}}^{\alpha}&=0\quad,\
    \end{aligned}
    &&
    \begin{aligned}
    \square B_{\mathsf{R}}^{\alpha}&=0\quad,\\\
    \partial_{\alpha} B_{\mathsf{R}}^{\alpha}&=0\quad.\
    \end{aligned}
\end{align}
Using \ref{LR} the $A$ and $B$ fields are expressed as 
\begin{align}\label{afield2}
    \begin{aligned}
    A_{\mathsf{L}}^{\alpha}&=\frac{1}{2}\left[(\closure{\Psi}+\closure{\Phi})\gamma^{\alpha}\frac{1-\gamma_{5}}{2}(\Psi+\Phi)-\frac{1}{m^{2}}\partial_{\lambda}(\closure{\Psi}+\closure{\Phi})\gamma^{\alpha}\frac{1-\gamma_{5}}{2}\partial^{\lambda}(\Psi+\Phi)\right]\quad,\\\
    A_{\mathsf{R}}^{\alpha}&=\frac{1}{2}\left[(\closure{\Psi}+\closure{\Phi})\gamma^{\alpha}\frac{1+\gamma_{5}}{2}(\Psi+\Phi)-\frac{1}{m^{2}}\partial_{\lambda}(\closure{\Psi}+\closure{\Phi})\gamma^{\alpha}\frac{1+\gamma_{5}}{2}\partial^{\lambda}(\Psi+\Phi)\right]\quad.\
    \end{aligned}
\end{align}

\begin{align}\label{afield3}
    \begin{aligned}
    B_{\mathsf{L}}^{\alpha}&=\frac{1}{2}\left[(\closure{\Psi}+\closure{\Phi})\gamma^{\alpha}\frac{1-\gamma_{5}}{2}(\Psi+\Phi)+\frac{1}{m^{2}}\partial_{\lambda}(\closure{\Psi}+\closure{\Phi})\gamma^{\alpha}\frac{1-\gamma_{5}}{2}\partial^{\lambda}(\Psi+\Phi)\right]\quad,\\\
    B_{\mathsf{R}}^{\alpha}&=\frac{1}{2}\left[(\closure{\Psi}+\closure{\Phi})\gamma^{\alpha}\frac{1+\gamma_{5}}{2}(\Psi+\Phi)+\frac{1}{m^{2}}\partial_{\lambda}(\closure{\Psi}+\closure{\Phi})\gamma^{\alpha}\frac{1+\gamma_{5}}{2}\partial^{\lambda}(\Psi+\Phi)\right]\quad.\
    \end{aligned}
\end{align}
For each solution of Dirac equation we compute the Maxwell fields $B^{\alpha}_{\mathsf{L,R}}$ fields and find that all are constant in spacetime and equal with $p^{\alpha}/\E$.
\begin{align}
  \begin{aligned}
B_{\mathsf{L,R}}^{\alpha}(\Psi_{1})=B_{\mathsf{L,R}}^{\alpha}(\Phi_{2})=B_{\mathsf{L,R}}^{\alpha}(\Psi_{2})=B_{\mathsf{L,R}}^{\alpha}(\Phi_{2})=\frac{p^{\alpha}}{\E}.
  \end{aligned}
  \end{align}
This result should be compared with Higgs Kibble cumbersome mechanism that postulates a scalar field $\phi$ which couples to a massless boson to obtain a factor in the wave equation for $W_{-}$ boson which play the same role as a mass term, and it is assumed that the scalar field is constant in space, while in our study we have proved, not postulated, the existence of a vector field, derived from Dirac equation, that is constant in spacetime. 
The Proca $A$ fields for $\Psi_{1}$ and $\Phi_{1}$ are equal
\begin{align}\label{proca1}
  \begin{aligned}
  &A_{\mathsf{L,R}}^{0}(\Psi_{1})=A_{\mathsf{L,R}}^{0}(\Phi_{1})=+\frac{m}{\E(\E+p_{3})}[p_{1}\cos(2px)-p_{2}\sin(2px)],\\
  &A_{\mathsf{L,R}}^{1}(\Psi_{1})=A_{\mathsf{L,R}}^{1}(\Phi_{1})=+\frac{m}{\E}\cos{(2px)},\\
  &A_{\mathsf{L,R}}^{2}(\Psi_{1})=A_{\mathsf{L,R}}^{2}(\Phi_{1})=-\frac{m}{\E}\sin{(2px)},\\
  &A_{\mathsf{L,R}}^{3}(\Psi_{1})=A_{\mathsf{L,R}}^{3}(\Phi_{1})=-\frac{m}{\E(\E+p_{3})}[p_{1}\cos(2px)-p_{2}\sin(2px)],
  \end{aligned}
  \end{align}
while Proca $A$ fields for $\Psi_{2}$ and $\Phi_{2}$ are equal up to sign
\begin{align}\label{proca2}
  \begin{aligned}
  &A_{\mathsf{L,R}}^{0}(\Psi_{2})=-A_{\mathsf{L,R}}^{0}(\Phi_{2})=+\frac{m}{\E(\E-p_{3})}[p_{1}\cos(2px)+p_{2}\sin(2px)],\\
  &A^{1}_{\mathsf{L,R}}(\Psi_{2})=-A_{\mathsf{L,R}}^{1}(\Phi_{2})=+\frac{m}{\E}\cos{(2px)},\\
  &A^{2}_{\mathsf{L,R}}(\Psi_{2})=-A_{\mathsf{L,R}}^{2}(\Phi_{2})=+\frac{m}{\E}\sin{(2px)},\\
  &A^{3}_{\mathsf{L,R}}(\Psi_{2})=-A^{3}_{\mathsf{L,R}}(\Phi_{2})=+\frac{m}{\E(\E-p_{3})}[p_{1}\cos(2px)+p_{2}\sin(2px)].
  \end{aligned}
  \end{align}
The four spinors $\Psi_{1,2}$ and $\Phi_{1,2}$ for a particle with mass $m$ generate two Proca potentials of mass $2m$ given by \ref{proca1} and \ref{proca2} further denoted by $S^{\alpha}_{1}$ and $S^{\alpha}_{2}$, defined by
  \begin{align}
  \begin{aligned}
  S^{\alpha}_{1}=\E A_{\mathsf{L}}^{\alpha}(\Psi_{1})=\E A_{\mathsf{L}}^{\alpha}(\Phi_{1}),
  \end{aligned}
  &&
  \begin{aligned}
  S^{\alpha}_{2}=\E A_{\mathsf{L}}^{\alpha}(\Psi_{2})=-\E A_{\mathsf{L}}^{\alpha}(\Phi_{2}),
  \end{aligned}
  \end{align}
which will be called spin fields due to the following relations of orthogonality to the four-momentum $p^{\alpha}=(\E, p_{1}, p_{2}, p_{3})$
\begin{align}
\begin{aligned}
p_{\alpha}S_{1}^{\alpha}=0,
\end{aligned}
&&
\begin{aligned}
p_{\alpha}S_{2}^{\alpha}=0.
\end{aligned}
\end{align}
The spin fields are solutions of the following 8 nonlinear, entangled partial differential equations which unifies Dirac and Proca equations
\begin{align}\label{unify}
\begin{aligned}
S^{\alpha}_{2}\partial_{\alpha}S^{\beta}_{1}=0\quad,
\end{aligned}
&&
\begin{aligned}
S^{\alpha}_{1}\partial_{\alpha}S^{\beta}_{2}=0\quad.
\end{aligned}
\end{align}
The spin fields $S_{1}$ and $S_{2}$ are entangled, the source of $S_{1}$ field is $S_{2}$ and vice versa, and the mass is not explicitly contained into the equations. The squared norm of the spin field is related to the mass, therefore the solutions $S_{1}$ and $S_{2}$ of coupled equations gives the mass spectrum
\begin{align}
\begin{aligned}
S_{1}^{\alpha}S^{1}_{\alpha}=-m^{2}\quad,
\end{aligned}&&
\begin{aligned}
S_{2}^{\alpha}S^{2}_{\alpha}=-m^{2}\quad.
\end{aligned}
\end{align}
\section{Algebra of Fundamental Discrete Symmetries}
The parity operator $\PP$ transform the spin fields $S_{1}$ and $S_{2}$ one into another
\begin{align}
\begin{aligned}
 \PP S^{1}(p,x)&= S^{2}(\PP p, \PP x)\quad,\\
\end{aligned}
&&
\begin{aligned}
 \PP S^{2}(p,x)&= S^{1}(\PP p, \PP x)\quad,\\
\end{aligned}
\end{align}
 in the same way that  charge operator $\C$ change left  and right particle spinors
\begin{align}
\begin{aligned}
X_{\mathsf{L}}(p,x)=\C X_{\mathsf{R}}(p,x)\quad,\
\end{aligned}
&&
\begin{aligned}
X_{\mathsf{R}}(p,x)=\C X_{\mathsf{L}}(p,x)\quad.\
\end{aligned}
\end{align}
Two commuting, antilinear and antiunitary operators $\PP\C=\TT$ and $\gamma_{5}\C=\ftilde{\TT}$
can be defined within the symmetry, which satisfy $\TT^{2}=\ftilde{\TT}^{2}=-1$, as required for time reversal operator and their action on spinors is
\begin{align}
\begin{aligned}
\TT\Phi_{1}(p,x)&=\gamma_{5}\Psi_{1}(p,x)\quad,\\
\ftilde{\TT}\Phi_{1}(p,x)&=\PP\Psi_{1}(p,x)\quad,
\end{aligned}&&
\begin{aligned}
\TT\Phi_{2}(p,x)&=i\PP\Psi_{2}(p,x)\quad,\\
\ftilde{\TT}\Phi_{2}(p,x)&=i\gamma_{5}\PP\Psi_{2}(p,x)\quad,
\end{aligned}
\end{align}
and on parity and chirality common eigenstates
\begin{align}
  \begin{aligned}
     \TT\Psi_{+\mathsf{L}}&=-\Phi_{-\mathsf{R}}\qquad,\\\
         \TT\Psi_{-\mathsf{L}}&=+ \Phi_{+\mathsf{R}}\qquad,\\\
          \TT\Psi_{+\mathsf{R}}&= -\Phi_{-\mathsf{L}}\qquad,\\\
           \TT\Psi_{-\mathsf{R}}&=+ \Phi_{+\mathsf{L}}\qquad,\
  \end{aligned}
  &&
   \begin{aligned}
     \ftilde{\TT}\Psi_{+\mathsf{L}}&=+\Phi_{-\mathsf{R}}\qquad,\\\
         \ftilde{\TT}\Psi_{-\mathsf{L}}&=+ \Phi_{+\mathsf{R}}\qquad,\\\
         \ftilde{\TT}\Psi_{+\mathsf{R}}&= -\Phi_{-\mathsf{L}}\qquad,\\\
           \ftilde{\TT}\Psi_{-\mathsf{R}}&=- \Phi_{+\mathsf{L}}\qquad,\
  \end{aligned}
  \end{align}
shows no linear combination with real coefficients of parity and chirality eigenstates is left invariant by action of $\TT$ and $\ftilde{\TT}$, therefore no conserved currents exists and no particles associated. The simultaneous application of charge conjugation, parity and $\TT$ or $\ftilde{\TT}$ is of fundamental importance 
\begin{align}
\begin{aligned}
\C\PP\TT=-1\quad,
\end{aligned}
&&
\begin{aligned}
\C\PP\ftilde{\TT}=+\gamma_{5}\PP\quad,
\end{aligned}
\end{align}
because the together with the expected $diag(-1,-1,-1,-1)$ we had the unexpected result $\gamma_{5}\PP=diag(1,-1,1,-1)$ is a consequence of charge conjugation, which change both chirality and parity.  The set of 8 operators \{$1$, $\gamma_{5}$, $\PP$, $\C$, $\gamma_{5}\PP$, $\gamma_{5}\C$, $\PP\C$ and $\gamma_{5}\PP\C$\} has 4 significant subsets listed below which span the algebra of split-quaternions, 
namely, linear combinations with real coefficients of four basis elements \{$1$, $e_{1}$, $e_{2}$, $e_{3}$\} that satisfy the following product rules: $e_{1}^2=-1$, $e_{2}^2=e_{3}^2=1$, $e_{1}e_{2}=e_{3}=-e_{2}e_{1}$ 
\begin{align}
\begin{aligned}
&\{1,\gamma_{5}\C, \C, \gamma{5}\},\\
&\{1, \gamma_{5}\C, -\PP,  \gamma_{5}\PP\C \},
\end{aligned}
&&
\begin{aligned}
&\{1,\PP\C, \C, \PP\},\\
&\{1, \PP\C,\gamma_{5}\PP\C, \gamma_{5}\}.
\end{aligned}
\end{align}
Changing the base to $K_{+}=\frac{1}{2}(e_{1}+e_{2})$, $K_{-}=\frac{1}{2}(e_{1}-e_{2})$ and $K_{3}=\frac{1}{2}e_{3}$ we get the algebra of $so(2,1)$
\begin{align}
\begin{aligned}
\left[K_{3},K_{+}\right]=+K_{+},
\end{aligned}
&&
\begin{aligned}
\left[K_{3},K_{-}\right]=-K_{-},
\end{aligned}
&&
\begin{aligned}
\left[K_{+},K_{-}\right]=-2K_{3}.
\end{aligned}
\end{align}
Using $\gamma_{5}$, $\PP$ and $\C$ we define a set of $6$ six  as follows
\begin{align}
\begin{aligned}
&u_{+}=\frac{1}{4}\left(\PP\C+\C+\gamma_{5}\C+\gamma_{5}\PP\C\right)&,\\
&u_{-}=\frac{1}{4}\left(\PP\C-\C+\gamma_{5}\C-\gamma_{5}\PP\C\right)&,\\
&u_{3}=\frac{1}{4}\left(\PP+\gamma_{5}\right)&,
\end{aligned}
&&
\begin{aligned}
&v_{+}=\frac{1}{4}\left(\PP\C+\C-\gamma_{5}\C-\gamma_{5}\PP\C\right)&,\\
&v_{-}=\frac{1}{4}\left(\PP\C-\C-\gamma_{5}\C+\gamma_{5}\PP\C\right)&,\\
&v_{3}=\frac{1}{4}\left(\PP-\gamma_{5}\right)&.
\end{aligned}
\end{align}
Both sets $u_{+}, u_{-}, u_{3}$ and $v_{+}, v_{-}, v_{3}$ satisfy $so(2,1)$ algebra
\begin{align}
\begin{aligned}
&\left[\,u_{\,3}\,,\,u_{+}\,\right]=\,u_{+}&,\\
&\left[\,u_{\,3}\,,\,u_{-}\,\right]=-u_{-}&,\\
&\left[\,u_{+}\,,\,u_{-}\,\right]=-2\,u_{3}&,\\
\end{aligned}
&&
\begin{aligned}
&\left[\,v_{\,3}\,,\,v_{+}\,\right]=\,v_{+}&,\\
&\left[\,v_{\,3}\,,\,v_{-}\,\right]=-v_{-}&,\\
&\left[\,v_{+}\,,\,v_{-}\,\right]=-2\,v_{3}&.\\
\end{aligned}
\end{align}
and all commutators brackets between sets $\{u_{+},u_{-},u_{3}\}$ and $\{v_{+},v_{-},v_{3}\}$ are zero, therefore the Lie algebra is decomposed in two $so(2,1)$ algebra, the adjoint representation is six dimensional. The algebra posses four anticomuting nilpotent operators $\PP \pm\gamma_{5}\C$ and  $\gamma_{5} \pm \PP\C$, therefore fermionic to extend naturally the algebra for the Yang-Mills theory
\begin{align}
\begin{aligned}
&a_{1}=\frac{1}{2}\left(\gamma_{5}-\PP\C\right)&,\\
&a_{2}=\frac{1}{2}\left(\PP+\gamma_{5}\C\right)&,\\
&b_{3}=\frac{1}{2}\C&,
\end{aligned}
&&
\begin{aligned}
&\overline{a}_{1}=\frac{1}{2}\left(\PP-\gamma_{5}\C\right)&,\\
&\overline{a}_{2}=\frac{1}{2}\left(\gamma_{5}+\PP\C\right)&,\\
&\overline{b}_{3}=\frac{1}{2}\gamma_{5}\PP\C&,
\end{aligned}
\end{align}
\begin{align}
\begin{aligned}
&a_{1}\,a_{1}=\overline{a}_{1}\,\overline{a}_{1}=0&,\\
&a_{1}\,\overline{a}_{1}=\overline{a}_{1}\,a_{1}=0&,
\end{aligned}&&
\begin{aligned}
&a_{2}\,a_{2}=\overline{a}_{2}\,\overline{a}_{2}=0\quad,\\
&a_{2}\,\overline{a}_{2}=\overline{a}_{2}\,a_{2}=0\quad,
\end{aligned}
\end{align}
together with where two commuting $ b_{3}$ and $ \overline{b}_{3} $ bosonic operators which are obtained from the fermionic operators
\begin{align}
\begin{aligned}
&b_{3}=\frac{1}{2}\left[\,a_{1}\,,\,a_{2}\,\right]=\frac{1}{2}\left[\,\overline{a}_{1}\,,\,\overline{a}_{2}\,\right]\quad,
\end{aligned}
&&
\begin{aligned}
&\overline{b}_{3}=\frac{1}{2}\left[\,a_{1}\,,\,\overline{a}_{2}\,\right]=\frac{1}{2}\left[\,\overline{a}_{1}\,,\,a_{2}\,\right]\quad,
\end{aligned}
\end{align}
and their Lie brackets with fermionic operators are given by
\begin{align}
\begin{aligned}
&\left[\,b_{3}\,,\,a_{1}\,\right]=\overline{a}_{1}&,\\
&\left[\,b_{3}\,,\,a_{2}\,\right]=-\overline{a}_{2}&,\\
&\left[\,b_{3}\,,\,\overline{a}_{1}\,\right]=a_{1}&,\\
&\left[\,b_{3}\,,\,\overline{a}_{2}\,\right]=-a_{2}&,\\
&\left[\,a_{1}\,,\,a_{2}\,\right]=2\,b_{3}&,\\
&\left[\,\overline{a}_{1}\,,\,\overline{a}_{2}\,\right]=2\,b_{3}&,
\end{aligned}
&&
\begin{aligned}
&\left[\,\overline{b}_{3}\,,\,a_{1}\,\right]=a_{1}&,\\
&\left[\,\overline{b}_{3}\,,\,a_{2}\,\right]=-a_{2}&,\\
&\left[\,\overline{b}_{3}\,,\,\overline{a}_{1}\,\right]=\overline{a}_{1}&,\\
&\left[\,\overline{b}_{3}\,,\,\overline{a}_{2}\,\right]=-\overline{a}_{2}&,\\
&\left[\,a_{1}\,,\,\overline{a}_{2}\,\right]=2\,\overline{b}_{3}&,\\
&\left[\,\overline{a}_{1}\,,\,a_{2}\,\right]=2\,\overline{b}_{3}&.
\end{aligned}
\end{align}
Now it is obvious that \{$a_{1}, a_{2}, b_{3}, \overline{a}_{1}, \overline{a}_{2},\overline{b}_{3} \}$ is a set from a larger algebra, with three $a$'s and three $b$'s. Adding a new pair of of nilpotent fermionic operators
$a_{3}$ and $\overline{a}_{3}$  we can define another two bosonic pairs $b_{1}$, $\overline{b}_{1}$ and $b_{2}$, $\overline{b}_{2}$ to get a 12-dimensional Lie algebra
\begin{align}
\begin{aligned}
&b_{1}=\frac{1}{2}\left[\,a_{2}\,,\,a_{3}\,\right]=\frac{1}{2}\left[\,\overline{a}_{2}\,,\,\overline{a}_{3}\,\right]&,\\
&b_{2}=\frac{1}{2}\left[\,a_{3}\,,\,a_{1}\,\right]=\frac{1}{2}\left[\,\overline{a}_{3}\,,\,\overline{a}_{1}\,\right]&,\\
&a_{3}\,a_{3}=\overline{a}_{3}\,\overline{a}_{3}=0&,\\
\end{aligned}
&&
\begin{aligned}
&\overline{b}_{1}=\frac{1}{2}\left[\,a_{2}\,,\,\overline{a}_{3}\,\right]=\frac{1}{2}\left[\,\overline{a}_{2}\,,\,a_{3}\,\right]&,\\
&\overline{b}_{2}=\frac{1}{2}\left[\,a_{3}\,,\,\overline{a}_{1}\,\right]=\frac{1}{2}\left[\,\overline{a}_{3}\,,\,a_{1}\,\right]&,\\
&a_{3}\,\overline{a}_{3}=\overline{a}_{3}\,a_{3}=0&.
\end{aligned}
\end{align}
We have two new sets of operators, namely $\{a_{2}, a_{3}, b_{1}, \overline{a}_{2}, \overline{a}_{3},\overline{b}_{1}\}$ and $\{a_{3}, a_{1}, b_{2}, \overline{a}_{3}, \overline{a}_{1},\overline{b}_{2} \}$ which form a sub-algebra in the same way that $\{ a_{1}, a_{2}, b_{3}, \overline{a}_{1}, \overline{a}_{2},\overline{b}_{3} \}$ does, their commutation relations are obtained by cyclic permutations of indices $\{1, 2, 3\}$,
\begin{align}
\begin{aligned}
&\left[\,b_{1}\,,\,a_{2}\,\right]=\overline{a}_{2}&,\\
&\left[\,b_{1}\,,\,a_{3}\,\right]=-\overline{a}_{3}&,\\
&\left[\,b_{1}\,,\,\overline{a}_{2}\,\right]=a_{2}&,\\
&\left[\,b_{1}\,,\,\overline{a}_{3}\,\right]=-a_{3}&,\\
&\left[\,a_{2}\,,\,a_{3}\,\right]=2\,b_{1}&,\\
&\left[\,\overline{a}_{2}\,,\,\overline{a}_{3}\,\right]=2\,b_{1}&,
\end{aligned}
&&
\begin{aligned}
&\left[\,\overline{b}_{1}\,,\,a_{2}\,\right]=a_{2}&,\\
&\left[\,\overline{b}_{1}\,,\,a_{3}\,\right]=-a_{3}&,\\
&\left[\,\overline{b}_{1}\,,\,\overline{a}_{2}\,\right]=\overline{a}_{2}&,\\
&\left[\,\overline{b}_{1}\,,\,\overline{a}_{3}\,\right]=-\overline{a}_{3}&,\\
&\left[\,a_{2}\,,\,\overline{a}_{3}\,\right]=2\,\overline{b}_{1}&,\\
&\left[\,\overline{a}_{2}\,,\,a_{3}\,\right]=2\,\overline{b}_{1}&.
\end{aligned}
\end{align}
and
\begin{align}
\begin{aligned}
&\left[\,b_{2}\,,\,a_{3}\,\right]=\overline{a}_{3}&,\\
&\left[\,b_{2}\,,\,a_{1}\,\right]=-\overline{a}_{1}&,\\
&\left[\,b_{2}\,,\,\overline{a}_{3}\,\right]=a_{3}&,\\
&\left[\,b_{2}\,,\,\overline{a}_{1}\,\right]=-a_{1}&,\\
&\left[\,a_{3}\,,\,a_{1}\,\right]=2\,b_{2}&,\\
&\left[\,\overline{a}_{3}\,,\,\overline{a}_{1}\,\right]=2\,b_{2}&,
\end{aligned}
&&
\begin{aligned}
&\left[\,\overline{b}_{2}\,,\,a_{3}\,\right]=a_{3}&,\\
&\left[\,\overline{b}_{2}\,,\,a_{1}\,\right]=-a_{1}&,\\
&\left[\,\overline{b}_{2}\,,\,\overline{a}_{3}\,\right]=\overline{a}_{3}&,\\
&\left[\,\overline{b}_{2}\,,\,\overline{a}_{1}\,\right]=-\overline{a}_{1}&,\\
&\left[\,a_{3}\,,\,\overline{a}_{1}\,\right]=2\,\overline{b}_{2}&,\\
&\left[\,\overline{a}_{3}\,,\,a_{1}\,\right]=2\,\overline{b}_{2}&.
\end{aligned}
\end{align}
The nilpotent fermionic operators satisfy the following relations 
\begin{align}
\begin{aligned}
&a_{1}\,\overline{a}_{1}=\overline{a}_{1}\,a_{1}=0&,\\
&a_{2}\,\overline{a}_{2}=\overline{a}_{2}\,a_{2}=0&,\\
&a_{1}\,\overline{a}_{2}=\overline{a}_{1}\,a_{2}&,\\
&a_{2}\,\overline{a}_{1}=\overline{a}_{2}\,a_{1}&,
\end{aligned}
&&
\begin{aligned}
&a_{1}\,a_{2}\,a_{1}=\overline{a}_{1}&,\\
&a_{2}\,a_{1}\,a_{2}=\overline{a}_{2}&,\\
&\overline{a}_{1}\,a_{2}\,a_{1}=a_{1}&,\\
&\overline{a}_{2}\,a_{1}\,a_{2}=a_{2}&,\\
\end{aligned}
&&
\begin{aligned}
&\overline{a}_{1}\,\overline{a}_{2}\,a_{1}=\overline{a}_{1}&,\\
&\overline{a}_{2}\,\overline{a}_{1}\,a_{2}=\overline{a}_{2}&,\\
&\overline{a}_{1}\,\overline{a}_{2}\,\overline{a}_{1}=a_{1}&,\\
&\overline{a}_{2}\,\overline{a}_{1}\,\overline{a}_{2}=a_{2}&,\\
\end{aligned}
\end{align}
and similar relations must hold for $\{\,a_{2}\,,\,a_{3}\,\}$ and $\{\,a_{3}\,,\,a_{1}\,\}$
\begin{align}
\begin{aligned}
&a_{k}\,\overline{a}_{k}=\overline{a}_{k}\,a_{k}=0&,\\
&a_{k}\,\overline{a}_{j}=\overline{a}_{k}\,a_{j}&,\\
\end{aligned}
&&
\begin{aligned}
&a_{k}\,a_{j}\,a_{k}=\overline{a}_{k}&,\\
&\overline{a}_{k}\,a_{j}\,a_{k}=a_{k}&,
\end{aligned}
&&
\begin{aligned}
&\overline{a}_{k}\,\overline{a}_{j}\,a_{k}=\overline{a}_{k}&,\\
&\overline{a}_{k}\,\overline{a}_{j}\,\overline{a}_{k}=a_{k}&
\end{aligned}
\end{align}
for $j,k=\{1,2,3\}$ and $j\neq k$.  Using this relations we get by direct calculus
\begin{align}
\begin{aligned}
&\left[\,b_{1}\,,\,b_{2}\,\right]=\left[\,\overline{b}_{1}\,,\,\overline{b}_{2}\,\right]=-\frac{1}{2}\,\overline{b}_{1}-\frac{1}{2}\,\overline{b}_{2}-\frac{1}{2}\,a_{3}\,b_{3}\,a_{3}&,\\
&\left[\,\overline{b}_{1}\,,\,b_{2}\,\right]=\left[\,b_{1}\,,\,\overline{b}_{2}\,\right]=-\frac{1}{2}\,b_{1}-\frac{1}{2}\,b_{2}-\frac{1}{2}\,a_{3}\,\overline{b}_{3}\,a_{3}&,\\
\end{aligned}
\end{align}
In order to cancel products like $\,\,a_{3}\,b_{3}\,a_{3}\,\,$ and $\,\,a_{3}\,\overline{b}_{3}\,a_{3}\,\,$we should require that either $a_{3}$ commutes with $b_{3}$ and $\overline{b}_{3}$ or any product of three $a$'s, like
\begin{align}
a_{i}\,a_{j}\,a_{k}\,,\, \overline{a}_{i}\,a_{j}\,a_{k}\,,\,\overline{a}_{i}\,\overline{a}_{j}\,a_{k}\,,\,
\overline{a}_{i}\,\overline{a}_{j}\,\overline{a}_{k}
\end{align} for $i\neq j \neq k\,$are invariat under cyclic permutations of $\,\{i,j, k\}\,$. Taking this into account we write down the rest of Lie algebra commutators
\begin{align}
\begin{aligned}
&\left[\,a_{k}\,,\,b_{k}\,\right]=0&,\\
&\left[\,\overline{a}_{k}\,,\,\overline{b}_{k}\,\right]=0&,
\end{aligned}
&&
\begin{aligned}
&\left[\,\overline{a}_{k}\,,\,b_{k}\,\right]=0&,\\
&\left[\,a_{k}\,,\,\overline{b}_{k}\,\right]=0&,\\
\end{aligned}
\end{align}
for $\,k=\{1,2,3\}\,$ and
\begin{align}
\begin{aligned}
&\left[\,b_{1}\,,\,b_{2}\,\right]=\left[\,\overline{b}_{1}\,,\,\overline{b}_{2}\,\right]=-\frac{1}{2}\,\overline{b}_{1}-\frac{1}{2}\,\overline{b}_{2}&,\\
&\left[\,b_{2}\,,\,b_{3}\,\right]=\left[\,\overline{b}_{2}\,,\,\overline{b}_{3}\,\right]=-\frac{1}{2}\,\overline{b}_{2}-\frac{1}{2}\,\overline{b}_{3}&,\\
&\left[\,b_{3}\,,\,b_{1}\,\right]=\left[\,\overline{b}_{3}\,,\,\overline{b}_{1}\,\right]=-\frac{1}{2}\,\overline{b}_{3}-\frac{1}{2}\,\overline{b}_{1}&,
\end{aligned}
&&
\begin{aligned}
&\left[\,\overline{b}_{1}\,,\,b_{2}\,\right]=\left[\,b_{1}\,,\,\overline{b}_{2}\,\right]=-\frac{1}{2}\,b_{1}-\frac{1}{2}\,b_{2}&,\\
&\left[\,\overline{b}_{2}\,,\,b_{3}\,\right]=\left[\,b_{2}\,,\,\overline{b}_{3}\,\right]=-\frac{1}{2}\,b_{2}-\frac{1}{2}\,b_{3}&,\\
&\left[\,\overline{b}_{3}\,,\,b_{1}\,\right]=\left[\,b_{3}\,,\,\overline{b}_{1}\,\right]=-\frac{1}{2}\,b_{3}-\frac{1}{2}\,b_{1}&.\\
\end{aligned}
\end{align}
In order to find the decomposition of the algebra of six operators $b_{1},b_{2}, b_{3},\overline{b}_{1}, \overline{b}_{2}, \overline{b}_{3}, $'s we define six new linear combinations
\begin{align}
\begin{aligned}
&u_{+}=b_{3}+\overline{b}_{3}+b_{1}+\overline{b}_{1}&,\\
&u_{-}=b_{3}+\overline{b}_{3}+b_{1}+\overline{b}_{1}&,\\
&u_{3}=b_{3}+\overline{b}_{3}&,
\end{aligned}
&&
\begin{aligned}
&v_{+}=b_{3}-\overline{b}_{3}+b_{1}-\overline{b}_{1}&,\\
&v_{-}=b_{3}-\overline{b}_{3}+b_{1}-\overline{b}_{1}&,\\
&v_{3}=b_{3}-\overline{b}_{3}&,
\end{aligned}
\end{align}
Both sets $u_{+}, u_{-}, u_{3}$ and $v_{+}, v_{-}, v_{3}$ satisfy $so(2,1)$ algebra
\begin{align}
\begin{aligned}
&\left[\,u_{\,3}\,,\,u_{+}\,\right]=\,u_{+}&,\\
&\left[\,u_{\,3}\,,\,u_{-}\,\right]=-u_{-}&,\\
&\left[\,u_{+}\,,\,u_{-}\,\right]=-2\,u_{3}&,\\
\end{aligned}
&&
\begin{aligned}
&\left[\,v_{\,3}\,,\,v_{+}\,\right]=\,v_{+}&,\\
&\left[\,v_{\,3}\,,\,v_{-}\,\right]=-v_{-}&,\\
&\left[\,v_{+}\,,\,v_{-}\,\right]=-2\,v_{3}&.\\
\end{aligned}
\end{align}
and all commutators brackets between sets $\{u_{+},u_{-},u_{3}\}$ and $\{v_{+},v_{-},v_{3}\}$ are zero, therefore the Lie algebra of $b$'s operators is decomposed in two $so(2,1)$ algebra.
\section{Conclusions}
This result, that Dirac equation is equivalent with two entangled Proca fields with mass $\mathsf{2m}$ and two constant massless fields, suggest that this representation unifies fermionic and bosonic fields, its algebra of 12 operators  could be used for study of weak and strong interactions. While the gauge symmetry $U(1)\times SU(2)\times SU(3)$ of the Standard Model is exact only when the particles are massless, the 12-dimensional Lie algebra proposed have the mass built in, i.e. it require neither cumbersome mechanism with arbitrary chosen fields and parameters nor spontaneous breaking symmetry to generate mass. In our model, the solutions of the equations $S^{\alpha}_{1}\partial_{\alpha}S^{\beta}_{2}=0$ and $S^{\alpha}_{2}\partial_{\alpha}S^{\beta}_{1}=0$ gives the mass of the field $S_{1}^{\alpha}S^{1}_{\alpha}=-m^{2}$ and $S_{2}^{\alpha}S^{2}_{\alpha}=-m^{2}$. The 12-dimensional Lie algebra of entangled spin can be applied to Yang-Mills theory together with a set of $12$ massless spin fields, six of them described by $S^{1}$ and six by $S^{2}$, their interaction to generate mass  will be the subject of a future study. 

\nocite{*} 
\printbibliography
\end{document}